\documentstyle[12pt]{article}
\def\frf#1{\vcenter{\hrule\hbox{\vrule\kern1pt
\vbox{\kern1pt\hbox{$\displaystyle#1$}%
\kern1pt}\kern1pt\vrule}\hrule}}

\def\dal{\frf{\phantom{o}}}
\def\bp{{\bf p}}
\def\bx{{\bf x}}
\def\sc{\scriptstyle}

\def\Sp{\rm{Sp}\,}

\def\newpic#1{%
\def\emline##1##2##3##4##5##6{%
\put(##1,##2){\special{em:point #1##3}}%
\put(##4,##5){\special{em:point #1##6}}%
\special{em:line #1##3,#1##6}}}
\newpic{}
\def\be{\begin{equation}}
\def\ee{\end{equation}}
\def\bea{\begin{eqnarray}}
\def\eea{\end{eqnarray}}
\def\text{\rm}
\def\be{\begin{equation}}
\def\ee{\end{equation}}
\def\bea{\begin{eqnarray}}
\def\eea{\end{eqnarray}}

\begin{document}

\begin{center}

\vskip 3cm
{\huge
Three-loop contributions \\
to the free energy of $\lambda\varphi^4$ QFT}
\vskip 1cm
{\large A.I.~Bugrij,
  V.N.~Shadura} \\
{\large  Bogoliubov Institute for Theoretical Physics,\\
252143 Kiev, Ukraine}\\
{e-mail: abugrij\@gluk.apc.org}

\end{center}
\vskip 1cm
\begin{abstract}
The massive scalar field with  $\lambda\varphi^4$ interaction placed
in $(3+1)$ dimensional box is considered. The sizes of the box are
$V\times \beta$ $(V=L^3$ is the volume, $T=1/\beta$ is the
temperature). The free energy is evaluated up to the 2nd order of
$PT$. The averaging on the vacuum fluctuations is separated from the
averaging on the thermal fluctuations explicitly. As result the
free-energy is expressed through the scattering amplitudes. We find
that in 3-loop approximation the expression for free energy coincides
with the ansatz of Bernstein, Dashen, Ma suggested on the base of
$S$-matrix formulation of statistical mechanics. The obtained
expressions are generalized for higher order of $PT$.
\end{abstract}

\section{Introduction}

The perturbation theory $(PT)$ for quantum field models at finite
temperature $T$ is elaborated enough [1]. Its formal distinction from
ordinary Feinman rules consists in that the loop summations on
discrete values of zero component of momentum appear (at $T\neq0$)
instead of the loop integrations (at $T=0$):
$$
T\sum_{p_0}\
\mathop{\to}_{T\to0}\
\frac{1}{2\pi}\int\limits_{-\infty}^{\infty}dp_0,
$$
where for the
boson fields
$$ p_0=2\pi lT,\quad l=0,\pm1,\pm2,\ldots
$$

The loop integral is treated as an averaging  on vacuum fluctuations.
It is clear that at $T\neq0$ both vacuum and thermal fluctuations of
the field take place. So, the difference between the loop sum and loop
integral is caused by thermal fluctuations alone. The standard task
of the quantum field theory (QFT) at finite temperature is the
calculation of radiative corrections to the free energy (or other
thermodynamic quantities) of the relativistic ideal gas as functions
of temperature, chemical potentials, renormalized masses, coupling
parameters, etc. There are no principal difficulties, but the job is
more cumbersome compared to usual Feinman diagrammar.

\newpage

\noindent
We want to
consider the problem from some other standpoint. We shall try to
separate the averaging on the thermal fluctuations from that of the
vacuum fluctuations explicitly and to represent the contribution
caused by interaction to the free energy as some integrals on
Bose-Einstein distribution of renormalized pure QFT values.

Note that probably the similar purpose was pursued in the papers [2,3]
devoted to the elaboration of QFT at finite temperature in real time
representation. But there was not achieved such simple and physically
qtransparent result as we obtained in the framework of standard
imaginary time representation.

We shall work out the calculations for the $\lambda\varphi^4$ QFT
model (extraction of the loop integrals from the corresponding sums,
renormalization) up to the 2nd order of PT, where the nontrivial
3-loops diagrams appear.

\section{The 1st order of $PT$}

For the beginning consider the first radiative correction to the free
energy. This exercise clarifies what result we would like to obtain.

The partition function of the scalar field placed to the thermostat of
the volume $V$ at the temperature $T=1/\beta$ is
$$
Z=\int{\cal{D}}\varphi\exp\bigl( -S_{0} \,[ \varphi]-S_I\,[
\varphi]\bigr),
$$
where the action in Euclidian metrics is
$$
S_0\,[\varphi]=\frac{1}{2}\int\limits_{\beta,V}d^4x\varphi(x)(m^2-
\frf{\phantom{o}})\varphi(x),
$$
$$
x=(x_0,{\bx}),\quad x^2=x_0^2+{\bx}^2,\quad \frf{\phantom{a}} =
{\partial^2\over\partial x^2_0}+(\nabla)^2,
$$
$$
S_I[\varphi]={\lambda\over4!}\int\limits_{\beta,V}d^4x\varphi^4(x).
$$

The free energy density in the 1st order on $\lambda$ is
$$
f(T)\equiv-{\ln Z\over \beta V}=f^{(0)}(T)+f^{(1)}(T),
$$
where at $V\to\infty$,
$$
f^{(0)}(T)={1\over2\beta
  V}\Sp\ln[\beta^2(m^2-\dal)]={1\over2(2\pi)^3}\int
  d^3{\bp}\left(T\sum_{p_0}\ln[\beta^2(m^2+p^2)]\right),
$$
$$
f^{(1)}(T)={1\over\beta V}\langle S_I[\varphi]\rangle={\lambda\over8}
\left({1\over(2\pi)^3}\int d^3{\bp}\left(T\sum_{p_0}{1\over
m^2+p^2}\right)\right)^2,
$$
$$
p=(p_0,\bp),\quad p^2=p_0^2+\bp^2
$$
Let us define the elementary diagrams to make the calculations more
compact and transparent as follows
\be
\newpic{1}
\unitlength=1.00mm
\special{em:linewidth 0.4pt}
\linethickness{0.4pt}
\begin{picture}(8.00,1.00)
\put(4.00,1.00){\circle{6.00}}
\end{picture}
\equiv-{1\over(2\pi)^3}\int d^3\bp\left(T\sum_{p_0}\ln[\beta^2
(m^2+p^2)]\right),
\ee
\be
\newpic{2}
\unitlength=1.00mm
\special{em:linewidth 0.4pt}
\linethickness{0.4pt}
\begin{picture}(8.00,1.00)
\put(4.00,1.00){\circle{6.00}}
\put(7.00,1.00){\circle*{1.00}}
\end{picture}
\equiv{1\over(2\pi)^3}\int d^3\bp\left(T\sum_{p_0}\frac{1}
{m^2+p^2}\right)=-{\partial\over\partial m^2}
\newpic{3}
\unitlength=1.00mm
\special{em:linewidth 0.4pt}
\linethickness{0.4pt}
\begin{picture}(8.00,1.00)
\put(4.00,1.00){\circle{6.00}}
\end{picture}.
\ee
It is easy to compute the sum on $p_0$ in (2)
\be
T\sum_{p_0}{1\over
p_0^2+\omega^2}={1\over2\pi}\int\limits_{-\infty}^{\infty} {dp_0\over
m^2+p^2}+{1\over\omega(e^{\beta\omega}-1)},\quad
\omega=\sqrt{m^2+\bp^2}.
\ee
One can see from this expression that
the averaging on vacuum and thermal fluctuations corresponding to one
loop sum can be represented additively. By integrating eq.(3) on
$m^2$ one obtains the temperature dependence of (1) i.e.
\be
f^{(0)}(T)=-{1\over2}
\newpic{4}
\unitlength=1.00mm
\special{em:linewidth 0.4pt}
\linethickness{0.4pt}
\begin{picture}(8.00,1.00)
\put(4.00,1.00){\circle{6.00}}
\end{picture}
={T\over(2\pi)^3}\int d^3p\ln(1-e^{-\beta\omega})+a+bT,
\ee
where $a$ and $b$ are irrelevant constants (energy density and pressure
of the vacuum), their values
 depend on regularization
scheme.  Let us denote the vacuum and thermal loops as
$$
\newpic{5}
\unitlength=1.00mm
\special{em:linewidth 0.4pt}
\linethickness{0.4pt} \begin{picture}(8.00,1.00)
\put(4.00,1.00){\circle{6.00}}
\put(4.00,1.00){\makebox(0,0)[cc]{$\sc T$}}
\end{picture}
=-{2T\over(2\pi)^3}\int d^3\bp\ln(1-e^{-\beta\omega}),
$$
$$
\newpic{6} \unitlength=1.00mm
\special{em:linewidth 0.4pt}
\linethickness{0.4pt}
\begin{picture}(8.00,1.00)
\put(4.00,1.00){\circle{6.00}}
\put(4.00,1.00){\makebox(0,0)[cc]{$\sc V$}}
\put(7.00,1.00){\circle*{1.00}}
\end{picture}
\equiv
\newpic{7}
\unitlength=1.00mm
\special{em:linewidth 0.4pt}
\linethickness{0.4pt}
\begin{picture}(8.00,1.00)
\put(4.00,1.00){\circle{6.00}}
\put(7.00,1.00){\circle*{1.00}}
\end{picture}
\bigg|_{T=0}={1\over(2\pi)^4}\int{d^4p\over m^2+p^2},
$$
\be
\newpic{8}
\unitlength=1.00mm
\special{em:linewidth 0.4pt}
\linethickness{0.4pt}
\begin{picture}(8.00,1.00)
\put(4.00,1.00){\circle{6.00}}
\put(4.00,1.00){\makebox(0,0)[cc]{$\sc T$}}
\put(7.00,1.00){\circle*{1.00}}
\end{picture}
\equiv
\newpic{9}
\unitlength=1.00mm
\special{em:linewidth 0.4pt}
\linethickness{0.4pt}
\begin{picture}(8.00,1.00)
\put(4.00,1.00){\circle{6.00}}
\put(7.00,1.00){\circle*{1.00}}
\end{picture}-
\newpic{10}
\unitlength=1.00mm
\special{em:linewidth 0.4pt}
\linethickness{0.4pt}
\begin{picture}(8.00,1.00)
\put(4.00,1.00){\circle{6.00}}
\put(4.00,1.00){\makebox(0,0)[cc]{V}}
\put(7.00,1.00){\circle*{1.00}}
\end{picture}
={2\over(2\pi)^3}\int{d^3\bp\over2\omega(e^{\beta\omega}-1)}.
\ee
In these notations the first order correction to the free energy has
the following diagrammatic representation
\be
f^{(1)}(T)={\lambda\over8}
\newpic{11}
\unitlength=1.00mm
\special{em:linewidth 0.4pt}
\linethickness{0.4pt}
\begin{picture}(14.00,1.00)
\put(4.00,1.00){\circle{6.00}}
\put(7.00,1.00){\circle*{1.00}}
\put(10.00,1.00){\circle{6.00}}
\end{picture}
={\lambda\over8}\bigl(
\newpic{12}
\unitlength=1.00mm
\special{em:linewidth 0.4pt}
\linethickness{0.4pt}
\begin{picture}(14.00,1.00)
\put(4.00,1.00){\circle{6.00}}
\put(4.00,1.00){\makebox(0,0)[cc]{$\sc T$}}
\put(10.00,1.00){\makebox(0,0)[cc]{$\sc T$}}
\put(7.00,1.00){\circle*{1.00}}
\put(10.00,1.00){\circle{6.00}}
\end{picture}
+2
\newpic{13}
\unitlength=1.00mm
\special{em:linewidth 0.4pt}
\linethickness{0.4pt}
\begin{picture}(14.00,1.00)
\put(4.00,1.00){\circle{6.00}}
\put(4.00,1.00){\makebox(0,0)[cc]{$\sc T$}}
\put(10.00,1.00){\makebox(0,0)[cc]{$\sc V$}}
\put(7.00,1.00){\circle*{1.00}}
\put(10.00,1.00){\circle{6.00}}
\end{picture}
+
\newpic{14}
\unitlength=1.00mm
\special{em:linewidth 0.4pt}
\linethickness{0.4pt}
\begin{picture}(14.00,1.00)
\put(4.00,1.00){\circle{6.00}}
\put(4.00,1.00){\makebox(0,0)[cc]{$\sc V$}}
\put(10.00,1.00){\makebox(0,0)[cc]{$\sc V$}}
\put(7.00,1.00){\circle*{1.00}}
\put(10.00,1.00){\circle{6.00}}
\end{picture}
\bigr).
\ee
Note, that
$
\newpic{15}
\unitlength=1.00mm
\special{em:linewidth 0.4pt}
\linethickness{0.4pt}
\begin{picture}(14.00,1.00)
\put(4.00,1.00){\circle{6.00}}
\put(7.00,1.00){\circle*{1.00}}
\put(10.00,1.00){\circle{6.00}}
\end{picture}=\bigl(
\newpic{16}
\unitlength=1.00mm
\special{em:linewidth 0.4pt}
\linethickness{0.4pt}
\begin{picture}(8.00,1.00)
\put(4.00,1.00){\circle{6.00}}
\put(7.00,1.00){\circle*{1.00}}
\end{picture}\bigr)^2$.
The divergent term
$
\newpic{17}
\unitlength=1.00mm
\special{em:linewidth 0.4pt}
\linethickness{0.4pt}
\begin{picture}(14.00,1.00)
\put(4.00,1.00){\circle{6.00}}
\put(4.00,1.00){\makebox(0,0)[cc]{$\sc V$}}
\put(10.00,1.00){\makebox(0,0)[cc]{$\sc V$}}
\put(7.00,1.00){\circle*{1.00}}
\put(10.00,1.00){\circle{6.00}}
\end{picture}
$
in (6) is irrelevant because it
does not depend on temperature. The first term in the r.h.s. of (6)
is convergent and it can be represented as the integral on
two-particles invariant phase volume with Bose-Einstein statistical
factors
\be
{\lambda\over8}
\newpic{18}
\unitlength=1.00mm
\special{em:linewidth 0.4pt}
\linethickness{0.4pt}
\begin{picture}(14.00,1.00)
\put(4.00,1.00){\circle{6.00}}
\put(4.00,1.00){\makebox(0,0)[cc]{$\sc T$}}
\put(10.00,1.00){\makebox(0,0)[cc]{$\sc T$}}
\put(7.00,1.00){\circle*{1.00}}
\put(10.00,1.00){\circle{6.00}}
\end{picture}
={1\over2!}{1\over(2\pi)^6}\int{d^3\bp_1d^3\bp_2\over2\omega_12\omega_2}
{\lambda\over(e^{\beta\omega_1}-1)(e^{\beta\omega_2}-1)}.
\ee
Keeping in mind that in the 1st order of $PT$ the $2\to 2$ scattering
amplitude is simply the coupling constant
$$
A^{(1)}_{2\to2}=-\lambda,
$$
one can consider (7) as the thermal averaging of the $2\to2$
scattering amplitude. Of course such interpretation looks artificial
enough. In the 2nd order of $PT$ the $2\to2$ scattering amplitude has
nontrivial dependence on momenta, also $3\to3$ amplitude appears. So,
the next order of $PT$ can bring more information about structure of
representations of type (7) . Before the 2nd order analyses
the diagram
$
\newpic{19}
\unitlength=1.00mm
\special{em:linewidth 0.4pt}
\linethickness{0.4pt} \vphantom{\biggl(\biggr)^2}
\begin{picture}(14.00,1.00) \put(4.00,1.00){\circle{6.00}}
\put(4.00,1.00){\makebox(0,0)[cc]{$\sc T$}}
\put(10.00,1.00){\makebox(0,0)[cc]{$\sc V$}}
\put(7.00,1.00){\circle*{1.00}}
\put(10.00,1.00){\circle{6.00}}
\end{picture}
$
have to be examined. It has nontrivial dependence on temperature and
is divergent. It is easy to see that this diagram can be removed by
the mass renormalization
$$
m^2\to m^2-\delta m^2_1.
$$
Then the zero order diagram in (4)
$$
\newpic{20}
\unitlength=1.00mm
\special{em:linewidth 0.4pt}
\linethickness{0.4pt}
\begin{picture}(8.00,1.00)
\put(4.00,1.00){\circle{6.00}}
\end{picture}
\to
\newpic{21}
\unitlength=1.00mm
\special{em:linewidth 0.4pt}
\linethickness{0.4pt}
\begin{picture}(8.00,1.00)
\put(4.00,1.00){\circle{6.00}}
\end{picture}
+ \delta m^2_1
\newpic{22}
\unitlength=1.00mm
\special{em:linewidth 0.4pt}
\linethickness{0.4pt}
\begin{picture}(8.00,1.00)
\put(4.00,1.00){\circle{6.00}}
\put(1.00,1.00){\circle*{1.00}}
\end{picture},
$$
and for the cancellation of the divergent temperature dependent term of
(6) is necessary $$ {\lambda\over4} \newpic{23}
\unitlength=1.00mm \special{em:linewidth 0.4pt} \linethickness{0.4pt}
\begin{picture}(14.00,1.00) \put(4.00,1.00){\circle{6.00}}
\put(4.00,1.00){\makebox(0,0)[cc]{$\sc V$}}
\put(10.00,1.00){\makebox(0,0)[cc]{$\sc T$}}
\put(7.00,1.00){\circle*{1.00}} \put(10.00,1.00){\circle{6.00}}
\end{picture} -{1\over2}\delta m^2_1
\newpic{24} \unitlength=1.00mm
\special{em:linewidth 0.4pt} \linethickness{0.4pt}
\begin{picture}(8.00,1.00) \put(4.00,1.00){\circle{6.00}}
\put(4.00,1.00){\makebox(0,0)[cc]{$\sc T$}}
\put(1.00,1.00){\circle*{1.00}}
\end{picture}
=0,
$$
that is
\be
\delta m^2_1={\lambda\over2}
\newpic{25}
\unitlength=1.00mm
\special{em:linewidth 0.4pt}
\linethickness{0.4pt}
\begin{picture}(8.00,1.00)
\put(4.00,1.00){\circle{6.00}}
\put(4.00,1.00){\makebox(0,0)[cc]{$\sc V$}}
\put(7.00,1.00){\circle*{1.00}}
\end{picture}
={\lambda\over2}{1\over(2\pi)^4}\int
{d^4p\over m^2+p^2}.
\ee

\section{The 2nd order of $PT$}

In the 2nd order of $PT$ the radiative correction to the free energy
is
\be
f^{(2)}(T)=-{1\over\beta V}\langle
S_I^2[\varphi]\rangle_c=-{\lambda^2\over48}\bigl(3
\newpic{26}
\unitlength=1.00mm
\special{em:linewidth 0.4pt}
\linethickness{0.4pt}
\begin{picture}(20.00,1.00)
\put(4.00,1.00){\circle{6.00}}
\put(16.00,1.00){\circle{6.00}}
\put(7.00,1.00){\circle*{1.00}}
\put(13.00,1.00){\circle*{1.00}}
\put(10.00,1.00){\circle{6.00}}
\end{picture}
+
\newpic{27}
\unitlength=1.00mm
\special{em:linewidth 0.4pt}
\linethickness{0.4pt}
\begin{picture}(12.00,1.00)
\put(4.00,1.00){\circle{6.00}}
\put(6.00,3.00){\circle*{1.00}}
\put(6.00,-1.00){\circle*{1.00}}
\put(8.00,1.00){\circle{6.00}}
\end{picture}
\bigr),
\ee
where the diagrams mean
\be
\newpic{28}
\unitlength=1.00mm
\special{em:linewidth 0.4pt}
\linethickness{0.4pt}
\begin{picture}(20.00,1.00)
\put(4.00,1.00){\circle{6.00}}
\put(16.00,1.00){\circle{6.00}}
\put(7.00,1.00){\circle*{1.00}}
\put(13.00,1.00){\circle*{1.00}}
\put(10.00,1.00){\circle{6.00}}
\end{picture}
=
\newpic{29}
\unitlength=1.00mm
\special{em:linewidth 0.4pt}
\linethickness{0.4pt}
\begin{picture}(8.00,1.00)
\put(4.00,1.00){\circle{6.00}}
\put(7.00,1.00){\circle*{1.00}}
\end{picture}
\times
\newpic{30}
\unitlength=1.00mm
\special{em:linewidth 0.4pt}
\linethickness{0.4pt}
\begin{picture}(8.00,1.00)
\put(4.00,1.00){\circle{6.00}}
\put(7.00,1.00){\circle*{1.00}}
\put(1.00,1.00){\circle*{1.00}}
\end{picture}
\times
\newpic{31}
\unitlength=1.00mm
\special{em:linewidth 0.4pt}
\linethickness{0.4pt}
\begin{picture}(8.00,1.00)
\put(4.00,1.00){\circle{6.00}}
\put(1.00,1.00){\circle*{1.00}}
\end{picture}
,
\ee
\be
\newpic{32}
\unitlength=1.00mm
\special{em:linewidth 0.4pt}
\linethickness{0.4pt}
\begin{picture}(8.00,1.00)
\put(4.00,1.00){\circle{6.00}}
\put(7.00,1.00){\circle*{1.00}}
\put(1.00,1.00){\circle*{1.00}}
\end{picture}
\equiv{1\over(2\pi)^3}\int
d^3\bp\left(T\sum_{p_0}{1\over(m^2+p^2)^2}\right)=
-{\partial\over\partial m^2}
\newpic{33}
\unitlength=1.00mm
\special{em:linewidth 0.4pt}
\linethickness{0.4pt}
\begin{picture}(8.00,1.00)
\put(4.00,1.00){\circle{6.00}}
\put(1.00,1.00){\circle*{1.00}}
\end{picture}
,
\ee
\bea
\newpic{34}
\unitlength=1.00mm
\special{em:linewidth 0.4pt}
\linethickness{0.4pt}
\begin{picture}(12.00,1.00)
\put(4.00,1.00){\circle{6.00}}
\put(6.00,3.00){\circle*{1.00}}
\put(6.00,-1.00){\circle*{1.00}}
\put(8.00,1.00){\circle{6.00}}
\end{picture}
&\equiv&{1\over(2\pi)^9}\int\left(\prod_{k=1}^{4}d^3\bp_k\right)\delta^3
(\bp_1+\bp_2+\bp_3+\bp_4)\times\nonumber\\
&\times&\left(T^3\sum_{p_{01}\ldots p_{04}}
{\delta(p_{01}+p_{02}+p_{03}+p_{04})\over(m^2+p^2_1)(m^2+p^2_2)
(m^2+p^2_3)(m^2+p^2_4)}\right),
\eea
where $\delta$-functions in (12) are mentioned as the Dirac one for
 integrals and the Kroneker symbol for sums.

The separation of  the vacuum and thermal loops from the diagram (10)
is not difficult due to eq.(5), but the corresponding procedure for
the diagram (12) requires some manipulations. Let us write the
diagram (12) as a sum of four possible combinations of vacuum and
thermal loops
\be
\newpic{35}
\unitlength=1.50mm
\special{em:linewidth 0.4pt}
\linethickness{0.4pt}
\begin{picture}(11.00,1.00)
\put(4.00,1.00){\circle{6.00}}
\put(5.50,3.50){\circle*{1.00}}
\put(5.50,-1.50){\circle*{1.00}}
\put(7.00,1.00){\circle{6.00}}
\end{picture}
=
\newpic{36}
\unitlength=1.50mm
\special{em:linewidth 0.4pt}
\linethickness{0.4pt}
\begin{picture}(11.00,1.00)
\put(4.00,1.00){\circle{6.00}}
\put(5.50,1.00){\makebox(0,0)[cc]{$\scriptstyle V\ V\ V$}}
\put(5.50,3.50){\circle*{1.00}}
\put(5.50,-1.50){\circle*{1.00}}
\put(7.00,1.00){\circle{6.00}}
\end{picture}
+4
\newpic{37}
\unitlength=1.50mm
\special{em:linewidth 0.4pt}
\linethickness{0.4pt}
\begin{picture}(11.00,1.00)
\put(4.00,1.00){\circle{6.00}}
\put(5.50,1.00){\makebox(0,0)[cc]{$\scriptstyle V\ T\ V$}}
\put(5.50,3.50){\circle*{1.00}}
\put(5.50,-1.50){\circle*{1.00}}
\put(7.00,1.00){\circle{6.00}}
\end{picture}
+6
\newpic{38}
\unitlength=1.50mm
\special{em:linewidth 0.4pt}
\linethickness{0.4pt}
\begin{picture}(11.00,1.00)
\put(4.00,1.00){\circle{6.00}}
\put(5.50,1.00){\makebox(0,0)[cc]{$\scriptstyle T\ V\ T$}}
\put(5.50,3.50){\circle*{1.00}}
\put(5.50,-1.50){\circle*{1.00}}
\put(7.00,1.00){\circle{6.00}}
\end{picture}
+4
\newpic{39}
\unitlength=1.50mm
\special{em:linewidth 0.4pt}
\linethickness{0.4pt}
\begin{picture}(11.00,1.00)
\put(4.00,1.00){\circle{6.00}}
\put(5.50,1.00){\makebox(0,0)[cc]{$\scriptstyle T\ T\ T$}}
\put(5.50,3.50){\circle*{1.00}}
\put(5.50,-1.50){\circle*{1.00}}
\put(7.00,1.00){\circle{6.00}}
\end{picture},
\ee

\medskip

\noindent
where the coefficients in the r.h.s. of (13) are chosen for the
combinatorial reason: the number of possibilities to cut one, two or
three lines connecting two vertices. The number of thermal loops in
diagrams (13) will be equal to the number of statistical factors
$(e^{\beta\omega}-1)^{-1}$ in corresponding integrals. Letus denote
the sum under the integral (12) by
$W(\omega_1,\omega_2,\omega_3,\omega_4)$
\be
W=T^3\sum_{p_{01}\ldots
p_{04}} {\delta(p_{01}+\cdots+p_{04})\over(p_{01}^2+\omega^2_1)\cdots
(p_{04}^2+\omega^2_4)},
\ee
and express the Kroneker $\delta$-symbol through the integral
$$
\delta(p_0)=T\int_{0}^{\beta}dx_0\, e^{ip_0x_0}.
$$
Then calculating the simple sum
$$
T\sum_{p_0}{e^{ip_0x_0}\over
p_0^2+\omega^2}={\cosh(x_0-\beta/2)\omega\over2\omega\cdot\sinh(\omega\beta/2)},
$$
we obtain for (14)
\be
W=\int\limits_{0}^{\beta}dx_0\prod_{k=1}^{4}\left(T\sum_{p_0}{e^{ip_0x_0}\over
p_0^2+\omega_k^2}\right)=\int\limits_{0}^{\beta}dx_0\prod_{k=1}^{4}
\left({\cosh(x_0-\beta/2)\omega_k\over2\omega_k\sinh\beta\omega_k/2}\right)
\ee
Direct integration in r.h.s. of (15) gives
\bea
W&=&2\left[\prod_{k=1}^{4}\bigl(2\omega_k(e^{\beta\omega_k}-1)\bigr)
\right]^{-1}
\cdot\left({e^{\beta(\omega_1+\omega_2+\omega_3+\omega_4)}-1\over
\omega_1+\omega_2+\omega_3+\omega_4}+\right.\nonumber\\
&+&4{e^{\beta(\omega_2+\omega_3+\omega_4)}\over -
\omega_1+\omega_2+\omega_3+\omega_4}+6
{e^{\beta(\omega_3+\omega_4)}\over -
\omega_1-\omega_2+\omega_3+\omega_4}+4
\left.{e^{\beta\omega_4}\over -
\omega_1-\omega_2-\omega_3+\omega_4}\right).
\eea
Despite the function $W$ is symmetric in its arguments by origin it is
more convenient to represent the result of integration of (15) in
asymmetric but compact form (16), accounting for the integral over
momenta in (12) pick out the symmetric part of $W$ only. Sorting (16)
accordingly to the number of statistical factors rewrite $W$ in the
following way
\bea
W=W_0+{4W_1\over2\omega_1(e^{\beta\omega_1}-1)}+{6W_2\over
2\omega_12\omega_2
(e^{\beta\omega_1}-1)(e^{\beta\omega_2}-1)}+\nonumber \\
{}+{4W_3\over
2\omega_12\omega_22\omega_3
(e^{\beta\omega_1}-1)(e^{\beta\omega_2}-1)(e^{\beta\omega_3}-1)},
\eea
where
\be
W_0=2\biggl(\prod_{k=1}^{4}2\omega_k\biggr)^{-1}{1\over
\omega_1+\omega_2+\omega_3+\omega_4 }={1\over(2\pi)^3}\int
\biggl(\prod_{k=1}^{4}{dp_{0k}\over
m^2+p^2}\biggr)\delta(p_{01}+\cdots+p_{04}),
\ee
\bea
W_1&=&2\biggl(\prod_{k=2}^{4}2\omega_k\biggr)^{-1}\left(
{1\over
\omega_1+\omega_2+\omega_3+\omega_4 }+
{1\over-\omega_1+\omega_2+\omega_3+\omega_4 }\right)=\nonumber\\
&=&{2\over(2\pi)^2}\int\biggl(\prod_{k=2}^{4}{dp_{0k}\over
m^2+p_k^2}\biggr)
\delta(p_{01}+\cdots+p_{04})\bigg|_{p_1^2=-m^2},
\eea
\bea
W_2&=&2\biggl(\prod_{k=3}^{4}2\omega_k\biggr)^{-1}\left(
{1\over\omega_1+\omega_2+\omega_3+\omega_4}+
{1\over -\omega_1-\omega_2+\omega_3+\omega_4 }+\right.\nonumber\\
&+&\left.{1\over -\omega_1+\omega_2+\omega_3+\omega_4 }+
{1\over \omega_1-\omega_2+\omega_3+\omega_4 }
\right)=
{2\over(2\pi)^2}\int\biggl(\prod_{k=2}^{4}{dp_{0k}\over
m^2+p_k^2}\biggr)\times\nonumber\\
&\times&\biggl(\delta(p_{01}+p_{02}+p_{03}+p_{04})+
\delta(-p_{01}+p_{02}+p_{03}+p_{04})\biggr)
\bigg|_{p_1^2=p_2^2=-m^2},
\eea
\bea
W_3&=&{1\over\omega_4}\left(
{1\over\omega_1+\omega_2+\omega_3+\omega_4}-
{1\over \omega_1+\omega_2+\omega_3-\omega_4 }+\right.\nonumber\\
&&\left.+{3\over -\omega_1+\omega_2+\omega_3+\omega_4 }-
{3\over -\omega_1+\omega_2+\omega_3-\omega_4 }
\right)=\nonumber\\
&=&2\left({1\over m^2+(p_1+p_2+p_3)^2}+{3\over m^2+(-p_1+p_2+p_3)^2}
\right)
\bigg|_{p_1^2=p_2^2=p_3^2=-m^2}.
\eea
Thus corresponding to the notation (13) and using (17)--(21) we
obtain
\bea
\newpic{40}
\unitlength=1.50mm
\special{em:linewidth 0.4pt}
\linethickness{0.4pt}
\begin{picture}(11.00,1.00)
\put(4.00,1.00){\circle{6.00}}
\put(5.50,1.00){\makebox(0,0)[cc]{$\scriptstyle V\ V\ V$}}
\put(5.50,3.50){\circle*{1.00}}
\put(5.50,-1.50){\circle*{1.00}}
\put(7.00,1.00){\circle{6.00}}
\end{picture}
={1\over(2\pi)^9}\int \left(\prod_{k=1}^{4}
d^3\bp_k\right)\delta^3(\bp_1+\bp_2+\bp_3+\bp_4)W_0=\nonumber\\
={1\over(2\pi)^{12}}\int \left(\prod_{k=1}^{4}
{d^4\bp_k\over m^2+p_k^2}\right)\delta^4(p_1+p_2+p_3+p_4),\nonumber
\eea
\bea
\newpic{41}
\unitlength=1.50mm
\special{em:linewidth 0.4pt}
\linethickness{0.4pt}
\begin{picture}(11.00,1.00)
\put(4.00,1.00){\circle{6.00}}
\put(5.50,1.00){\makebox(0,0)[cc]{$\scriptstyle V\ T\ V$}}
\put(5.50,3.50){\circle*{1.00}}
\put(5.50,-1.50){\circle*{1.00}}
\put(7.00,1.00){\circle{6.00}}
\end{picture}
&=&{2\over(2\pi)^3}\int{d^3\bp_1\over2\omega_1(e^{\beta\omega}-1)}\left[
{1\over(2\pi)^8}\int\left(\prod_{k=2}^{4}
{d^4\bp_k\over m^2+p_k^2}\right)\delta^4(p_1+\cdots+p_4)\right],\\
&&{\rm at}\  p_1^2=-m^2.\nonumber
\eea
The expression in the square brackets of r.h.s. (22) is none other
than ordinary two-loops contribution to the inverse propagator, with
the external momentum being on the mass shell:  \be
\biggl[\cdots\biggr]= \newpic{42}
\unitlength=1.00mm \special{em:linewidth 0.4pt}
\linethickness{0.4pt} \begin{picture}(13.00,1.00)(-2.00,0.50)
\put(4.00,1.00){\circle{6.00}} \put(1.00,1.00){\circle*{1.00}}
\emline{-2.00}{1.00}{1}{10.00}{1.00}{2}
\put(7.00,1.00){\circle*{1.00}}
\end{picture}
{}_{\displaystyle p_1^2=-m^2}\ .
\ee
Further
\bea
\newpic{43}
\unitlength=1.50mm
\special{em:linewidth 0.4pt}
\linethickness{0.4pt}
\begin{picture}(11.00,1.00)
\put(4.00,1.00){\circle{6.00}}
\put(5.50,1.00){\makebox(0,0)[cc]{$\scriptstyle T\ V\ T$}}
\put(5.50,3.50){\circle*{1.00}}
\put(5.50,-1.50){\circle*{1.00}}
\put(7.00,1.00){\circle{6.00}}
\end{picture}
&=&{2\over(2\pi)^6}\int\left(\prod_{k=1}^{2}{d^3\bp_k\over
2\omega_k(e^{\beta\omega_k}-1)}\right)\times\nonumber\\
&\times&{\rm Re}\left[{1\over(2\pi)^4}\int {d^4p\over
m^2+p^2}\biggl({1\over m^2+(p+p_1+p_2)^2}+{1\over
m^2+(p-p_1+p_2)^2}\biggr)\right],\\
&&\qquad\qquad{\rm at} \ p_1^2=p_2^2=-m^2.\nonumber
\eea
Here in the square brackets of r.h.s. (24) one can  recognize one-loop
contribution ($s$- and
$u$-channel terms) to the two-particle forward scattering
amplitude on the mass shell
\be
\biggl[\cdots\biggr]=
\newpic{44}
\unitlength=1.50mm \special{em:linewidth 0.6pt}
\linethickness{0.6pt} \begin{picture}(50.00,1.00)(35,40)
\put(42.00,38.00){\oval(6.00,6.00)[t]}
\put(42.00,42.00){\oval(6.00,6.00)[b]}
\put(51.00,40.00){\makebox(0,0)[cc]{$+$}}
\put(39.50,40.00){\circle*{0.80}}
\put(44.50,40.00){\circle*{0.80}}
\put(56.50,40.00){\oval(5.00,6.00)[r]}
\put(60.50,40.00){\oval(5.00,4.00)[l]}
\put(58.83,41.50){\circle*{0.80}}
\put(58.83,38.33){\circle*{0.80}}
\emline{60.52}{42.01}{1}{61.26}{41.75}{2}
\emline{60.52}{38.00}{3}{61.33}{38.35}{4}
\emline{61.26}{41.73}{5}{65.10}{38.00}{6}
\emline{61.34}{38.36}{7}{62.66}{39.67}{8}
\emline{62.66}{39.67}{9}{62.80}{39.62}{10}
\emline{62.80}{39.62}{11}{63.00}{39.58}{12}
\emline{63.00}{39.58}{13}{63.16}{39.61}{14}
\emline{63.16}{39.61}{15}{63.30}{39.69}{16}
\emline{63.30}{39.69}{17}{63.45}{39.85}{18}
\emline{63.45}{39.85}{19}{63.49}{39.98}{20}
\emline{63.49}{39.98}{21}{63.45}{40.17}{22}
\emline{63.45}{40.17}{23}{63.41}{40.30}{24}
\emline{63.41}{40.30}{25}{65.12}{42.00}{26}
\put(37.50,43.00){\makebox(0,0)[cc]{$p_2$}}
\put(37.50,37.00){\makebox(0,0)[cc]{$p_1$}}
\put(47.50,43.00){\makebox(0,0)[cc]{$p_2$}}
\put(47.50,37.00){\makebox(0,0)[cc]{$p_1$}}
\put(55.00,43.00){\makebox(0,0)[cc]{$p_2$}}
\put(55.00,37.00){\makebox(0,0)[cc]{$p_1$}}
\put(67.00,43.00){\makebox(0,0)[cc]{$p_2$}}
\put(67.00,37.00){\makebox(0,0)[cc]{$p_1$}}
\end{picture}
\ee

\medskip\noindent
Just the second ($u$-channel) term testifies zero angle scattering.
Note that $t$-channel term at zero angle coincides with the diagram
(11) in our notations

\medskip
\be
\newpic{45}
\unitlength=1.50mm
\special{em:linewidth 0.6pt} \linethickness{0.6pt}
\begin{picture}(10.00,1.00)(2,-1) \put(7.00,0.00){\oval(6.00,6.00)[l]}
\put(3.00,0.00){\oval(6.00,6.00)[r]} \put(5.00,2.17){\circle*{1.00}}
\put(5.00,-2.33){\circle*{1.00}}
\put(1.00,3.00){\makebox(0,0)[cc]{$p_2$}}
\put(1.00,-3.00){\makebox(0,0)[cc]{$p_1$}}
\put(9.00,3.00){\makebox(0,0)[cc]{$p_2$}}
\put(9.00,-3.00){\makebox(0,0)[cc]{$p_1$}}
\end{picture}=
\newpic{46}
\unitlength=1.00mm
\special{em:linewidth 0.4pt}
\linethickness{0.4pt}
\begin{picture}(8.00,1.00)
\put(4.00,1.00){\circle{6.00}}
\put(4.00,1.00){\makebox(0,0)[cc]{$\sc V$}}
\put(1.00,1.00){\circle*{1.00}}
\put(7.00,1.00){\circle*{1.00}}
\end{picture},
\ee

\medskip\noindent
We shall find it in the diagram (10).
For the last term of r.h.s. (13) we have
\bea
\newpic{47}
\unitlength=1.50mm
\special{em:linewidth 0.4pt}
\linethickness{0.4pt}
\begin{picture}(11.00,1.00)
\put(4.00,1.00){\circle{6.00}}
\put(5.50,1.00){\makebox(0,0)[cc]{$\scriptstyle T\ T\ T$}}
\put(5.50,3.50){\circle*{1.00}}
\put(5.50,-1.50){\circle*{1.00}}
\put(7.00,1.00){\circle{6.00}}
\end{picture}
={1\over(2\pi)^9}\int
\left(\prod_{k=1}^{3}{d^3\bp_k\over
2\omega_k(e^{\beta\omega_k}-1)}\right)
\left[{1\over m^2+(p_1+p_2+p_3)^2}\right.+\nonumber\\
+\left.{3\over m^2+(-p_1+p_2+p_3)^2}\right],\quad {\rm at}\
p_1^2=p_2^2=p_3^2=-m^2.
\eea
It is easy to see allowing for the momenta permutation symmetry that
here the square brackets expression represents the contribution of four
diagrams to the $3\to3$ zero-angle scattering amplitude

\medskip
\be
\biggl[\cdots\biggr]=
\newpic{48}
\unitlength=0.50mm
\special{em:linewidth 0.6pt}
\linethickness{0.6pt}
\begin{picture}(46.00,1.00)(-4,-1)
\emline{5.00}{0.00}{1}{35.00}{0.00}{2}
\emline{15.00}{0.00}{3}{5.00}{10.00}{4}
\emline{15.00}{0.00}{5}{5.00}{-10.00}{6}
\put(15.00,0.00){\circle*{2.00}}
\emline{25.00}{0.00}{7}{35.00}{10.00}{8}
\emline{25.00}{0.00}{9}{35.00}{-10.00}{10}
\put(25.00,0.00){\circle*{2.00}}
\put(4.00,10.00){\makebox(0,0)[rc]{$p_3$}}
\put(4.00,0.00){\makebox(0,0)[rc]{$p_2$}}
\put(4.00,-10.00){\makebox(0,0)[rc]{$p_1$}}
\put(36.00,10.00){\makebox(0,0)[lc]{$p_3$}}
\put(36.00,0.00){\makebox(0,0)[lc]{$p_2$}}
\put(36.00,-10.00){\makebox(0,0)[lc]{$p_1$}}
\end{picture}+
\newpic{49}
\unitlength=0.50mm
\special{em:linewidth 0.6pt}
\linethickness{0.6pt}
\begin{picture}(46.00,1.00)(-4,11)
\emline{20.00}{19.50}{1}{35.00}{19.50}{2}
\emline{20.00}{19.50}{3}{5.00}{24.50}{4}
\emline{20.00}{19.50}{5}{5.00}{14.50}{6}
\emline{5.00}{4.50}{7}{20.00}{4.50}{8}
\emline{20.00}{4.50}{9}{35.00}{9.50}{10}
\emline{20.00}{4.50}{11}{35.00}{-0.50}{12}
\emline{20.00}{19.50}{13}{20.00}{4.50}{14}
\put(20.00,4.50){\circle*{2.00}}
\put(20.00,19.50){\circle*{2.00}}
\put(4.00,24.50){\makebox(0,0)[rc]{$p_3$}}
\put(4.00,14.50){\makebox(0,0)[rc]{$p_2$}}
\put(4.00,4.50){\makebox(0,0)[rc]{$p_1$}}
\put(36.00,19.50){\makebox(0,0)[lc]{$p_1$}}
\put(36.00,9.50){\makebox(0,0)[lc]{$p_2$}}
\put(36.00,-0.50){\makebox(0,0)[lc]{$p_3$}}
\end{picture}+
\newpic{50}
\unitlength=0.50mm
\special{em:linewidth 0.6pt}
\linethickness{0.6pt}
\begin{picture}(46.00,1.00)(-4,11)
\emline{20.00}{19.50}{1}{35.00}{19.50}{2}
\emline{20.00}{19.50}{3}{5.00}{24.50}{4}
\emline{20.00}{19.50}{5}{5.00}{14.50}{6}
\emline{5.00}{4.50}{7}{20.00}{4.50}{8}
\emline{20.00}{4.50}{9}{35.00}{9.50}{10}
\emline{20.00}{4.50}{11}{35.00}{-0.50}{12}
\emline{20.00}{19.50}{13}{20.00}{4.50}{14}
\put(20.00,4.50){\circle*{2.00}}
\put(20.00,19.50){\circle*{2.00}}
\put(4.00,24.50){\makebox(0,0)[rc]{$p_3$}}
\put(4.00,14.50){\makebox(0,0)[rc]{$p_1$}}
\put(4.00,4.50){\makebox(0,0)[rc]{$p_2$}}
\put(36.00,19.50){\makebox(0,0)[lc]{$p_2$}}
\put(36.00,9.50){\makebox(0,0)[lc]{$p_1$}}
\put(36.00,-0.50){\makebox(0,0)[lc]{$p_3$}}
\end{picture}+
\newpic{51}
\unitlength=0.50mm
\special{em:linewidth 0.6pt}
\linethickness{0.6pt}
\begin{picture}(46.00,1.00)(-4,11)
\emline{20.00}{19.50}{1}{35.00}{19.50}{2}
\emline{20.00}{19.50}{3}{5.00}{24.50}{4}
\emline{20.00}{19.50}{5}{5.00}{14.50}{6}
\emline{5.00}{4.50}{7}{20.00}{4.50}{8}
\emline{20.00}{4.50}{9}{35.00}{9.50}{10}
\emline{20.00}{4.50}{11}{35.00}{-0.50}{12}
\emline{20.00}{19.50}{13}{20.00}{4.50}{14}
\put(20.00,4.50){\circle*{2.00}}
\put(20.00,19.50){\circle*{2.00}}
\put(4.00,24.50){\makebox(0,0)[rc]{$p_1$}}
\put(4.00,14.50){\makebox(0,0)[rc]{$p_2$}}
\put(4.00,4.50){\makebox(0,0)[rc]{$p_3$}}
\put(36.00,19.50){\makebox(0,0)[lc]{$p_3$}}
\put(36.00,9.50){\makebox(0,0)[lc]{$p_2$}}
\put(36.00,-0.50){\makebox(0,0)[lc]{$p_1$}}
\end{picture}
\ .
\ee

\bigskip\noindent
So, we make shure that the 3-loop contribution to the free energy also
exhibits the structure of Bose-Einstein averaging of the scattering
amplitudes.

\section{Renormalization}

At the same time we see that the temperature dependent expressions
(22), (24) are divergent owing to divergences of the vacuum loops in
(23), (25), (26). Show how these divergences are absorbed due to
mass and coupling constant renormalization
\be
m^2\to m^2-\delta m_1^2-\delta m_2^2,\quad \lambda\to
\lambda+\delta\lambda.
\ee
The first order correction $\delta m_1^2$ is already fixed by (8).
For $\delta m_2^2$ and $\delta \lambda$ put
\be
\delta\lambda=a_1\lambda^2,\quad \delta m_2^2=a_2\lambda^2.
\ee
This leads to
$$
\newpic{52}
\unitlength=1.00mm
\special{em:linewidth 0.4pt} \linethickness{0.4pt}
\begin{picture}(8.00,1.00) \put(4.00,1.00){\circle{6.00}}
\end{picture} \to
\newpic{53}
\unitlength=1.00mm
\special{em:linewidth 0.4pt}
\linethickness{0.4pt} \begin{picture}(8.00,1.00)
\put(4.00,1.00){\circle{6.00}} \end{picture}
+{\lambda\over2}
\newpic{54}
\unitlength=1.00mm
\special{em:linewidth 0.4pt} \linethickness{0.4pt}
\begin{picture}(14.00,1.00) \put(4.00,1.00){\circle{6.00}}
\put(4.00,1.00){\makebox(0,0)[cc]{$\sc V$}}
\put(7.00,1.00){\circle*{1.00}}
\put(10.00,1.00){\circle{6.00}} \end{picture}+
{\lambda^2\over8}
\newpic{55}
\unitlength=1.00mm
\special{em:linewidth 0.4pt} \linethickness{0.4pt}
\begin{picture}(20.00,1.00) \put(4.00,1.00){\circle{6.00}}
\put(4.00,1.00){\makebox(0,0)[cc]{$\sc V$}}
\put(16.00,1.00){\makebox(0,0)[cc]{$\sc V$}}
\put(16.00,1.00){\circle{6.00}}
\put(7.00,1.00){\circle*{1.00}}
\put(13.00,1.00){\circle*{1.00}}
\put(10.00,1.00){\circle{6.00}}
\end{picture}
+{\lambda}^2 a_2
\newpic{56}
\unitlength=1.00mm
\special{em:linewidth 0.4pt}
\linethickness{0.4pt}
\begin{picture}(8.00,1.00)
\put(4.00,1.00){\circle{6.00}}
\put(1.00,1.00){\circle*{1.00}}
\end{picture},
$$
$$
\newpic{57}
\unitlength=1.00mm
\special{em:linewidth 0.4pt}
\linethickness{0.4pt}
\begin{picture}(14.00,1.00)
\put(4.00,1.00){\circle{6.00}}
\put(7.00,1.00){\circle*{1.00}}
\put(10.00,1.00){\circle{6.00}}
\end{picture}\to
\newpic{58}
\unitlength=1.00mm
\special{em:linewidth 0.4pt}
\linethickness{0.4pt}
\begin{picture}(14.00,1.00)
\put(4.00,1.00){\circle{6.00}}
\put(7.00,1.00){\circle*{1.00}}
\put(10.00,1.00){\circle{6.00}}
\end{picture}+\lambda
\newpic{59}
\unitlength=1.00mm
\special{em:linewidth 0.4pt}
\linethickness{0.4pt}
\begin{picture}(20.00,1.00)
\put(4.00,1.00){\circle{6.00}}
\put(4.00,1.00){\makebox(0,0)[cc]{$\sc V$}}
\put(16.00,1.00){\circle{6.00}}
\put(7.00,1.00){\circle*{1.00}}
\put(13.00,1.00){\circle*{1.00}}
\put(10.00,1.00){\circle{6.00}}
\end{picture}
+\lambda a_1
\newpic{60}
\unitlength=1.00mm
\special{em:linewidth 0.4pt}
\linethickness{0.4pt}
\begin{picture}(14.00,1.00)
\put(4.00,1.00){\circle{6.00}}
\put(7.00,1.00){\circle*{1.00}}
\put(10.00,1.00){\circle{6.00}}
\end{picture},
$$

\medskip\noindent
and for the second radiative correction (9) to the free energy we
find
\bea
f^{(2)}(T)=&-&{\lambda^2\over48}\bigl(3
\newpic{61}
\unitlength=1.00mm
\special{em:linewidth 0.4pt}
\linethickness{0.4pt}
\begin{picture}(20.00,1.00)
\put(4.00,1.00){\circle{6.00}}
\put(16.00,1.00){\circle{6.00}}
\put(7.00,1.00){\circle*{1.00}}
\put(13.00,1.00){\circle*{1.00}}
\put(10.00,1.00){\circle{6.00}}
\end{picture}
+
\newpic{62}
\unitlength=1.00mm
\special{em:linewidth 0.4pt}
\linethickness{0.4pt}
\begin{picture}(11.00,1.00)
\put(4.00,1.00){\circle{6.00}}
\put(5.50,3.50){\circle*{1.00}}
\put(5.50,-1.50){\circle*{1.00}}
\put(7.00,1.00){\circle{6.00}}
\end{picture}
+3
\newpic{63}
\unitlength=1.00mm
\special{em:linewidth 0.4pt}
\linethickness{0.4pt}
\begin{picture}(20.00,1.00)
\put(4.00,1.00){\circle{6.00}}
\put(4.00,1.00){\makebox(0,0)[cc]{$\sc V$}}
\put(16.00,1.00){\makebox(0,0)[cc]{$\sc V$}}
\put(16.00,1.00){\circle{6.00}}
\put(7.00,1.00){\circle*{1.00}}
\put(13.00,1.00){\circle*{1.00}}
\put(10.00,1.00){\circle{6.00}}
\end{picture}-\nonumber\\
&-&6
\newpic{64}
\unitlength=1.00mm
\special{em:linewidth 0.4pt}
\linethickness{0.4pt}
\begin{picture}(20.00,1.00)
\put(4.00,1.00){\circle{6.00}}
\put(4.00,1.00){\makebox(0,0)[cc]{$\sc V$}}
\put(16.00,1.00){\circle{6.00}}
\put(7.00,1.00){\circle*{1.00}}
\put(13.00,1.00){\circle*{1.00}}
\put(10.00,1.00){\circle{6.00}}
\end{picture}
-6a_1
\newpic{65}
\unitlength=1.00mm
\special{em:linewidth 0.4pt}
\linethickness{0.4pt}
\begin{picture}(14.00,1.00)
\put(4.00,1.00){\circle{6.00}}
\put(7.00,1.00){\circle*{1.00}}
\put(10.00,1.00){\circle{6.00}}
\end{picture}
+24a_2
\newpic{66}
\unitlength=1.00mm
\special{em:linewidth 0.4pt}
\linethickness{0.4pt}
\begin{picture}(8.00,1.00)
\put(4.00,1.00){\circle{6.00}}
\put(1.00,1.00){\circle*{1.00}}
\end{picture}
\bigr).
\eea
Writing separately the contributions to $f^{(2)}(T)$ corresponding
to different numbers of thermal loops we obtain:

\underbar{3 $T$-loops}
\be
g_3=3
\newpic{67}
\unitlength=1.00mm
\special{em:linewidth 0.4pt}
\linethickness{0.4pt}
\begin{picture}(20.00,1.00)
\put(4.00,1.00){\circle{6.00}}
\put(4.00,1.00){\makebox(0,0)[cc]{$\sc T$}}
\put(16.00,1.00){\makebox(0,0)[cc]{$\sc T$}}
\put(10.00,1.00){\makebox(0,0)[cc]{$\sc T$}}
\put(16.00,1.00){\circle{6.00}}
\put(7.00,1.00){\circle*{1.00}}
\put(13.00,1.00){\circle*{1.00}}
\put(10.00,1.00){\circle{6.00}}
\end{picture}
+4
\newpic{68}
\unitlength=1.50mm
\special{em:linewidth 0.4pt}
\linethickness{0.4pt}
\begin{picture}(11.00,1.00)
\put(4.00,1.00){\circle{6.00}}
\put(5.50,1.00){\makebox(0,0)[cc]{$\scriptstyle T\ T\ T$}}
\put(5.50,3.50){\circle*{1.00}}
\put(5.50,-1.50){\circle*{1.00}}
\put(7.00,1.00){\circle{6.00}}
\end{picture}.
\ee

\underbar{2 $T$-loops}
\be
g_2=9
\newpic{69}
\unitlength=1.00mm
\special{em:linewidth 0.4pt}
\linethickness{0.4pt}
\begin{picture}(20.00,1.00)
\put(4.00,1.00){\circle{6.00}}
\put(4.00,1.00){\makebox(0,0)[cc]{$\sc T$}}
\put(10.00,1.00){\makebox(0,0)[cc]{$\sc V$}}
\put(16.00,1.00){\makebox(0,0)[cc]{$\sc T$}}
\put(16.00,1.00){\circle{6.00}}
\put(7.00,1.00){\circle*{1.00}}
\put(13.00,1.00){\circle*{1.00}}
\put(10.00,1.00){\circle{6.00}}
\end{picture}
-6a_1
\newpic{70}
\unitlength=1.00mm
\special{em:linewidth 0.4pt}
\linethickness{0.4pt}
\begin{picture}(14.00,1.00)
\put(4.00,1.00){\circle{6.00}}
\put(4.00,1.00){\makebox(0,0)[cc]{$\sc T$}}
\put(10.00,1.00){\makebox(0,0)[cc]{$\sc T$}}
\put(7.00,1.00){\circle*{1.00}}
\put(10.00,1.00){\circle{6.00}}
\end{picture}
+6\biggl(
\newpic{71}
\unitlength=1.50mm
\special{em:linewidth 0.4pt}
\linethickness{0.4pt}
\begin{picture}(11.00,1.00)
\put(4.00,1.00){\circle{6.00}}
\put(5.50,1.00){\makebox(0,0)[cc]{$\scriptstyle T\ V\ T$}}
\put(5.50,3.50){\circle*{1.00}}
\put(5.50,-1.50){\circle*{1.00}}
\put(7.00,1.00){\circle{6.00}}
\end{picture}
-
\newpic{72}
\unitlength=1.00mm
\special{em:linewidth 0.4pt}
\linethickness{0.4pt}
\begin{picture}(20.00,1.00)
\put(4.00,1.00){\circle{6.00}}
\put(4.00,1.00){\makebox(0,0)[cc]{$\sc T$}}
\put(10.00,1.00){\makebox(0,0)[cc]{$\sc V$}}
\put(16.00,1.00){\makebox(0,0)[cc]{$\sc T$}}
\put(16.00,1.00){\circle{6.00}}
\put(7.00,1.00){\circle*{1.00}}
\put(13.00,1.00){\circle*{1.00}}
\put(10.00,1.00){\circle{6.00}}
\end{picture}
\biggr).\ee
The expression in brackets in r.h.s. (33) is finite
\bea
\newpic{73}
\unitlength=1.50mm
\special{em:linewidth 0.4pt}
\linethickness{0.4pt}
\begin{picture}(11.00,1.00)
\put(4.00,1.00){\circle{6.00}}
\put(5.50,1.00){\makebox(0,0)[cc]{$\scriptstyle T\ V\ T$}}
\put(5.50,3.50){\circle*{1.00}}
\put(5.50,-1.50){\circle*{1.00}}
\put(7.00,1.00){\circle{6.00}}
\end{picture}
-
\newpic{74}
\unitlength=1.00mm
\special{em:linewidth 0.4pt}
\linethickness{0.4pt}
\begin{picture}(20.00,1.00)
\put(4.00,1.00){\circle{6.00}}
\put(4.00,1.00){\makebox(0,0)[cc]{$\sc T$}}
\put(10.00,1.00){\makebox(0,0)[cc]{$\sc V$}}
\put(16.00,1.00){\makebox(0,0)[cc]{$\sc T$}}
\put(16.00,1.00){\circle{6.00}}
\put(7.00,1.00){\circle*{1.00}}
\put(13.00,1.00){\circle*{1.00}}
\put(10.00,1.00){\circle{6.00}}
\end{picture}
={2\over(2\pi)^6}\int{d^3\bp_1d^3\bp_2\over2\omega_12\omega_2
(e^{\beta\omega_1}-1)(e^{\beta\omega_2}-1)}\times\nonumber\\
\times
{\rm Re}\, \left[{1\over(2\pi)^4}\int{d^4p\over
m^2+p^2}\biggl({1\over m^2+(p+p_1+p_2)^2}+{1\over m^2+(p-p_1+p_2)^2}-
{2\over m^2+p^2}\biggr)\right],
\eea
therefore $g_2$ will be finite under condition of cancellation first
two terms in r.h.s. (33) that is
\be
a_1={3\over2}
\newpic{75}
\unitlength=1.00mm
\special{em:linewidth 0.4pt}
\linethickness{0.4pt}
\begin{picture}(8.00,1.00)
\put(4.00,1.00){\circle{6.00}}
\put(4.00,1.00){\makebox(0,0)[cc]{$\sc V$}}
\put(7.00,1.00){\circle*{1.00}}
\put(1.00,1.00){\circle*{1.00}}
\end{picture}
\ee
With allowance for (35) we have for

\underbar{1 $T$-loop}
\be
g_1=4
\newpic{76}
\unitlength=1.50mm
\special{em:linewidth 0.4pt}
\linethickness{0.4pt}
\begin{picture}(11.00,1.00)
\put(4.00,1.00){\circle{6.00}}
\put(5.50,1.00){\makebox(0,0)[cc]{$\scriptstyle V\ T\ V$}}
\put(5.50,3.50){\circle*{1.00}}
\put(5.50,-1.50){\circle*{1.00}}
\put(7.00,1.00){\circle{6.00}}
\end{picture}
-18
\newpic{77}
\unitlength=1.00mm
\special{em:linewidth 0.4pt}
\linethickness{0.4pt}
\begin{picture}(20.00,1.00)
\put(4.00,1.00){\circle{6.00}}
\put(4.00,1.00){\makebox(0,0)[cc]{$\sc V$}}
\put(10.00,1.00){\makebox(0,0)[cc]{$\sc T$}}
\put(16.00,1.00){\makebox(0,0)[cc]{$\sc V$}}
\put(16.00,1.00){\circle{6.00}}
\put(7.00,1.00){\circle*{1.00}}
\put(13.00,1.00){\circle*{1.00}}
\put(10.00,1.00){\circle{6.00}}
\end{picture}
+24a_2
\newpic{78}
\unitlength=1.00mm
\special{em:linewidth 0.4pt}
\linethickness{0.4pt}
\begin{picture}(8.00,1.00)
\put(4.00,1.00){\makebox(0,0)[cc]{$\sc T$}}
\put(4.00,1.00){\circle{6.00}}
\put(1.00,1.00){\circle*{1.00}}
\end{picture}.
\ee

\medskip\noindent
Using (22), (23) for the diagram
$
\vphantom{\biggm|^2}
\newpic{79}
\unitlength=1.50mm
\special{em:linewidth 0.4pt}
\linethickness{0.4pt}
\begin{picture}(11.00,1.00)
\put(4.00,1.00){\circle{6.00}}
\put(5.50,1.00){\makebox(0,0)[cc]{$\scriptstyle V\ T\ V$}}
\put(5.50,3.50){\circle*{1.00}}
\put(5.50,-1.50){\circle*{1.00}}
\put(7.00,1.00){\circle{6.00}}
\end{picture}
$ one can see from (36) that $g_1$ vanishes
if
\be
a_2={3\over4}
\newpic{80}
\unitlength=1.00mm
\special{em:linewidth 0.4pt}
\linethickness{0.4pt}
\begin{picture}(14.00,1.00)
\put(4.00,1.00){\circle{6.00}}
\put(4.00,1.00){\makebox(0,0)[cc]{$\sc T$}}
\put(10.00,1.00){\makebox(0,0)[cc]{$\sc T$}}
\put(7.00,1.00){\circle*{1.00}}
\put(13.00,1.00){\circle*{1.00}}
\put(10.00,1.00){\circle{6.00}}
\end{picture}
-{1\over6}
\newpic{81}
\unitlength=1.00mm
\special{em:linewidth 0.4pt}
\linethickness{0.4pt}
\begin{picture}(13.00,1.00)(-2.00,0.50)
\put(4.00,1.00){\circle{6.00}}
\put(1.00,1.00){\circle*{1.00}}
\emline{-2.00}{1.00}{1}{10.00}{1.00}{2}
\put(7.00,1.00){\circle*{1.00}}
\end{picture}
{}_{\displaystyle p^2=-m^2}.
\ee
In usual diagram language [4] the obtained mass and coupling constant
renormalization (8), (29), (30), (35), (37) looks as follows
\bea
\delta m_1^2&=&{\lambda\over2}
\newpic{82}
\unitlength=1.00mm
\special{em:linewidth 0.4pt}
\linethickness{0.4pt}
\begin{picture}(13.00,1.00)(-2.00,0.50)
\put(4.00,4.00){\circle{6.00}}
\put(4.00,1.00){\circle*{1.00}}
\emline{-2.00}{1.00}{1}{10.00}{1.00}{2}
\end{picture}
,\quad \delta
m_2^2={3\lambda^2\over4}
\newpic{83}
\unitlength=1.00mm
\special{em:linewidth 0.4pt}
\linethickness{0.4pt}
\begin{picture}(13.00,10.00)(-2.00,0.50)
\put(4.00,4.00){\circle{6.00}}
\put(4.00,10.00){\circle{6.00}}
\put(4.00,7.00){\circle*{1.00}}
\put(4.00,1.00){\circle*{1.00}}
\emline{-2.00}{1.00}{1}{10.00}{1.00}{2}
\end{picture}
-{\lambda\over6}
\newpic{84}
\unitlength=1.00mm
\special{em:linewidth 0.4pt}
\linethickness{0.4pt}
\begin{picture}(13.00,1.00)(-2.00,0.50)
\put(4.00,1.00){\circle{6.00}}
\put(1.00,1.00){\circle*{1.00}}
\emline{-2.00}{1.00}{1}{10.00}{1.00}{2}
\put(7.00,1.00){\circle*{1.00}}
\end{picture}
{}_{\displaystyle p^2=-m^2}
\ ,\nonumber\\
\delta\lambda&=&{3\lambda^2\over2}
\newpic{85}
\unitlength=1.50mm
\special{em:linewidth 0.6pt}
\linethickness{0.6pt}
\begin{picture}(12.00,5.00)(37,39.50)
\put(42.00,38.00){\oval(6.00,6.00)[t]}
\put(42.00,42.00){\oval(6.00,6.00)[b]}
\put(39.50,40.00){\circle*{0.80}}
\put(44.50,40.00){\circle*{0.80}}
\end{picture}
 \quad {\rm at}\ p_i=0.
\eea
One can see from (38) that the renormalization obtained here from
the condition of convergency temperature dependent part of free energy
is just the same that follows from the condition of 1PI Green's
functions be finite
\bea
z\Gamma^{(2)}(p^2)=0\quad&{\rm at}&
p^2=-m^2,\nonumber\\
\Gamma^{(4)}(p_i)=-\lambda\quad&{\rm at}& p_i=0.\nonumber
\eea
Finally accounting for (31--34) the renormalized free energy in the 2nd
order of $PT$ is

\bea f(T)&=&-{1\over2}
\newpic{86} \unitlength=1.00mm
\special{em:linewidth 0.4pt} \linethickness{0.4pt}
\begin{picture}(8.00,1.00) \put(4.00,1.00){\makebox(0,0)[cc]{$\sc T$}}
\put(4.00,1.00){\circle{6.00}} \end{picture}
+{\lambda\over8} \newpic{87}
\unitlength=1.00mm \special{em:linewidth 0.4pt}
\linethickness{0.4pt} \begin{picture}(14.00,1.00)
\put(10.00,1.00){\makebox(0,0)[cc]{$\sc T$}}
\put(4.00,1.00){\makebox(0,0)[cc]{$\sc T$}}
\put(10.00,1.00){\circle{6.00}}
\put(4.00,1.00){\circle{6.00}}
\put(7.00,1.00){\circle*{1.00}}
\end{picture}
-\nonumber\\
&-&{\lambda^2\over48}\biggl[3
\newpic{88}
\unitlength=1.00mm
\special{em:linewidth 0.4pt}
\linethickness{0.4pt}
\begin{picture}(20.00,1.00)
\put(10.00,1.00){\makebox(0,0)[cc]{$\sc T$}}
\put(16.00,1.00){\makebox(0,0)[cc]{$\sc T$}}
\put(4.00,1.00){\makebox(0,0)[cc]{$\sc T$}}
\put(16.00,1.00){\circle{6.00}}
\put(10.00,1.00){\circle{6.00}}
\put(4.00,1.00){\circle{6.00}}
\put(7.00,1.00){\circle*{1.00}}
\put(13.00,1.00){\circle*{1.00}}
\end{picture}
+4
\newpic{89}
\unitlength=1.50mm
\special{em:linewidth 0.4pt}
\linethickness{0.4pt}
\begin{picture}(11.00,1.00)
\put(4.00,1.00){\circle{6.00}}
\put(5.50,1.00){\makebox(0,0)[cc]{$\scriptstyle T\ T\ T$}}
\put(5.50,3.50){\circle*{1.00}}
\put(5.50,-1.50){\circle*{1.00}}
\put(7.00,1.00){\circle{6.00}}
\end{picture}
+6\biggl(
\newpic{90}
\unitlength=1.50mm
\special{em:linewidth 0.4pt}
\linethickness{0.4pt}
\begin{picture}(11.00,1.00)
\put(4.00,1.00){\circle{6.00}}
\put(5.50,1.00){\makebox(0,0)[cc]{$\scriptstyle T\ V\ T$}}
\put(5.50,3.50){\circle*{1.00}}
\put(5.50,-1.50){\circle*{1.00}}
\put(7.00,1.00){\circle{6.00}}
\end{picture}
-
\newpic{91}
\unitlength=1.00mm
\special{em:linewidth 0.4pt}
\linethickness{0.4pt}
\begin{picture}(20.00,1.00)
\put(10.00,1.00){\makebox(0,0)[cc]{$\sc V$}}
\put(16.00,1.00){\makebox(0,0)[cc]{$\sc T$}}
\put(4.00,1.00){\makebox(0,0)[cc]{$\sc T$}}
\put(16.00,1.00){\circle{6.00}}
\put(10.00,1.00){\circle{6.00}}
\put(4.00,1.00){\circle{6.00}}
\put(7.00,1.00){\circle*{1.00}}
\put(13.00,1.00){\circle*{1.00}}
\end{picture}
\biggr)\biggr].
\eea

\section{Thermal averaging}

Let the contributions corresponding to averaging on 2- and
3-particles phase volume be denoted through $f_2(T)$ and $f_3(T)$. Then
from (39) we obtain \be f_2(T)={\lambda\over8} \newpic{92}
\unitlength=1.00mm \special{em:linewidth 0.4pt} \linethickness{0.4pt}
\begin{picture}(14.00,1.00) \put(10.00,1.00){\makebox(0,0)[cc]{$\sc
T$}} \put(4.00,1.00){\makebox(0,0)[cc]{$\sc T$}}
\put(10.00,1.00){\circle{6.00}} \put(4.00,1.00){\circle{6.00}}
\put(7.00,1.00){\circle*{1.00}} \end{picture}
-{\lambda^2\over8}\biggl( \newpic{93}
\unitlength=1.50mm \special{em:linewidth 0.4pt}
\linethickness{0.4pt} \begin{picture}(11.00,1.00)
\put(4.00,1.00){\circle{6.00}}
\put(5.50,1.00){\makebox(0,0)[cc]{$\scriptstyle T\ V\ T$}}
\put(5.50,3.50){\circle*{1.00}}
\put(5.50,-1.50){\circle*{1.00}}
\put(7.00,1.00){\circle{6.00}}
\end{picture}
-
\newpic{94}
\unitlength=1.00mm
\special{em:linewidth 0.4pt}
\linethickness{0.4pt}
\begin{picture}(20.00,1.00)
\put(10.00,1.00){\makebox(0,0)[cc]{$\sc V$}}
\put(16.00,1.00){\makebox(0,0)[cc]{$\sc T$}}
\put(4.00,1.00){\makebox(0,0)[cc]{$\sc T$}}
\put(16.00,1.00){\circle{6.00}}
\put(10.00,1.00){\circle{6.00}}
\put(4.00,1.00){\circle{6.00}}
\put(7.00,1.00){\circle*{1.00}}
\put(13.00,1.00){\circle*{1.00}}
\end{picture}
\biggr),
\ee
\be
f_3(T)=-{\lambda^2\over48}\biggl(3
\newpic{95}
\unitlength=1.00mm
\special{em:linewidth 0.4pt}
\linethickness{0.4pt}
\begin{picture}(20.00,1.00)
\put(10.00,1.00){\makebox(0,0)[cc]{$\sc T$}}
\put(16.00,1.00){\makebox(0,0)[cc]{$\sc T$}}
\put(4.00,1.00){\makebox(0,0)[cc]{$\sc T$}}
\put(16.00,1.00){\circle{6.00}}
\put(10.00,1.00){\circle{6.00}}
\put(4.00,1.00){\circle{6.00}}
\put(7.00,1.00){\circle*{1.00}}
\put(13.00,1.00){\circle*{1.00}}
\end{picture}
+4
\newpic{96}
\unitlength=1.50mm
\special{em:linewidth 0.4pt}
\linethickness{0.4pt}
\begin{picture}(11.00,1.00)
\put(4.00,1.00){\circle{6.00}}
\put(5.50,1.00){\makebox(0,0)[cc]{$\scriptstyle T\ T\ T$}}
\put(5.50,3.50){\circle*{1.00}}
\put(5.50,-1.50){\circle*{1.00}}
\put(7.00,1.00){\circle{6.00}}
\end{picture}
\biggr),
\ee
and with allowance for the analytic expressions of diagrams (5),
(7), (27), (34) we have for (40), (41)
\be
f_2(T)=-{1\over2!}{1\over(2\pi)^6}\int
\left(\prod_{k=1}^{2}{d^3\bp_k\over2\omega_k(e^{\beta\omega_k}-1)}\right)
{\rm Re}\,\bigl[A_{2\to2}(p_1,p_2;p_1,p_2)\bigr],
\ee
\be
f_3(T)=-{1\over3!}{1\over(2\pi)^9}\int
\left(\prod_{k=1}^{3}{d^3\bp_k\over2\omega_k(e^{\beta\omega_k}-1)}\right)
{\rm Re}\,\bigl[A_{3\to3}(p_1,p_2,p_3;p_1,p_2,p_3)\bigr].
\ee
Some comments are to be made
concerning the representation of
$f_3(t)$. As it follows from (27), (28) the diagram $
\vphantom{\biggm|^2}
\newpic{97}
\unitlength=1.50mm
\special{em:linewidth 0.4pt}
\linethickness{0.4pt}
\begin{picture}(11.00,1.00)
\put(4.00,1.00){\circle{6.00}}
\put(5.50,1.00){\makebox(0,0)[cc]{$\scriptstyle T\ T\ T$}}
\put(5.50,3.50){\circle*{1.00}}
\put(5.50,-1.50){\circle*{1.00}}
\put(7.00,1.00){\circle{6.00}}
\end{picture}
$ contains
only 4 from 10 possible tree diagrams which saturate the $3\to3$
amplitude in the 2nd order of $PT$. The other 6 diagrams, i.e.

\medskip
\be
\newpic{98}
\unitlength=0.50mm
\special{em:linewidth 0.6pt}
\linethickness{0.6pt}
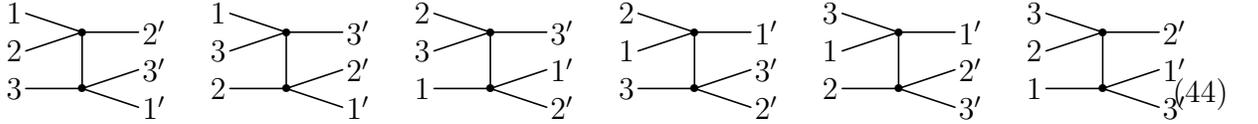
\begin{picture}(46.00,1.00)(-4,11)
\emline{20.00}{19.50}{1}{35.00}{19.50}{2}
\emline{20.00}{19.50}{3}{5.00}{24.50}{4}
\emline{20.00}{19.50}{5}{5.00}{14.50}{6}
\emline{5.00}{4.50}{7}{20.00}{4.50}{8}
\emline{20.00}{4.50}{9}{35.00}{9.50}{10}
\emline{20.00}{4.50}{11}{35.00}{-0.50}{12}
\emline{20.00}{19.50}{13}{20.00}{4.50}{14}
\put(20.00,4.50){\circle*{2.00}}
\put(20.00,19.50){\circle*{2.00}}
\put(4.00,24.50){\makebox(0,0)[rc]{$1$}}
\put(4.00,14.50){\makebox(0,0)[rc]{$2$}}
\put(4.00,4.50){\makebox(0,0)[rc]{$3$}}
\put(36.00,19.50){\makebox(0,0)[lc]{$2'$}}
\put(36.00,9.50){\makebox(0,0)[lc]{$3'$}}
\put(36.00,-0.50){\makebox(0,0)[lc]{$1'$}}
\end{picture}
\quad
\newpic{99}
\unitlength=0.50mm
\special{em:linewidth 0.6pt}
\linethickness{0.6pt}
\begin{picture}(46.00,1.00)(-4,11)
\emline{20.00}{19.50}{1}{35.00}{19.50}{2}
\emline{20.00}{19.50}{3}{5.00}{24.50}{4}
\emline{20.00}{19.50}{5}{5.00}{14.50}{6}
\emline{5.00}{4.50}{7}{20.00}{4.50}{8}
\emline{20.00}{4.50}{9}{35.00}{9.50}{10}
\emline{20.00}{4.50}{11}{35.00}{-0.50}{12}
\emline{20.00}{19.50}{13}{20.00}{4.50}{14}
\put(20.00,4.50){\circle*{2.00}}
\put(20.00,19.50){\circle*{2.00}}
\put(4.00,24.50){\makebox(0,0)[rc]{$1$}}
\put(4.00,14.50){\makebox(0,0)[rc]{$3$}}
\put(4.00,4.50){\makebox(0,0)[rc]{$2$}}
\put(36.00,19.50){\makebox(0,0)[lc]{$3'$}}
\put(36.00,9.50){\makebox(0,0)[lc]{$2'$}}
\put(36.00,-0.50){\makebox(0,0)[lc]{$1'$}}
\end{picture}\quad
\newpic{100}
\unitlength=0.50mm
\special{em:linewidth 0.6pt}
\linethickness{0.6pt}
\begin{picture}(46.00,1.00)(-4,11)
\emline{20.00}{19.50}{1}{35.00}{19.50}{2}
\emline{20.00}{19.50}{3}{5.00}{24.50}{4}
\emline{20.00}{19.50}{5}{5.00}{14.50}{6}
\emline{5.00}{4.50}{7}{20.00}{4.50}{8}
\emline{20.00}{4.50}{9}{35.00}{9.50}{10}
\emline{20.00}{4.50}{11}{35.00}{-0.50}{12}
\emline{20.00}{19.50}{13}{20.00}{4.50}{14}
\put(20.00,4.50){\circle*{2.00}}
\put(20.00,19.50){\circle*{2.00}}
\put(4.00,24.50){\makebox(0,0)[rc]{$2$}}
\put(4.00,14.50){\makebox(0,0)[rc]{$3$}}
\put(4.00,4.50){\makebox(0,0)[rc]{$1$}}
\put(36.00,19.50){\makebox(0,0)[lc]{$3'$}}
\put(36.00,9.50){\makebox(0,0)[lc]{$1'$}}
\put(36.00,-0.50){\makebox(0,0)[lc]{$2'$}}
\end{picture}\quad
\newpic{101}
\unitlength=0.50mm
\special{em:linewidth 0.6pt}
\linethickness{0.6pt}
\begin{picture}(46.00,1.00)(-4,11)
\emline{20.00}{19.50}{1}{35.00}{19.50}{2}
\emline{20.00}{19.50}{3}{5.00}{24.50}{4}
\emline{20.00}{19.50}{5}{5.00}{14.50}{6}
\emline{5.00}{4.50}{7}{20.00}{4.50}{8}
\emline{20.00}{4.50}{9}{35.00}{9.50}{10}
\emline{20.00}{4.50}{11}{35.00}{-0.50}{12}
\emline{20.00}{19.50}{13}{20.00}{4.50}{14}
\put(20.00,4.50){\circle*{2.00}}
\put(20.00,19.50){\circle*{2.00}}
\put(4.00,24.50){\makebox(0,0)[rc]{$2$}}
\put(4.00,14.50){\makebox(0,0)[rc]{$1$}}
\put(4.00,4.50){\makebox(0,0)[rc]{$3$}}
\put(36.00,19.50){\makebox(0,0)[lc]{$1'$}}
\put(36.00,9.50){\makebox(0,0)[lc]{$3'$}}
\put(36.00,-0.50){\makebox(0,0)[lc]{$2'$}}
\end{picture}\quad
\newpic{102}
\unitlength=0.50mm
\special{em:linewidth 0.6pt}
\linethickness{0.6pt}
\begin{picture}(46.00,1.00)(-4,11)
\emline{20.00}{19.50}{1}{35.00}{19.50}{2}
\emline{20.00}{19.50}{3}{5.00}{24.50}{4}
\emline{20.00}{19.50}{5}{5.00}{14.50}{6}
\emline{5.00}{4.50}{7}{20.00}{4.50}{8}
\emline{20.00}{4.50}{9}{35.00}{9.50}{10}
\emline{20.00}{4.50}{11}{35.00}{-0.50}{12}
\emline{20.00}{19.50}{13}{20.00}{4.50}{14}
\put(20.00,4.50){\circle*{2.00}}
\put(20.00,19.50){\circle*{2.00}}
\put(4.00,24.50){\makebox(0,0)[rc]{$3$}}
\put(4.00,14.50){\makebox(0,0)[rc]{$1$}}
\put(4.00,4.50){\makebox(0,0)[rc]{$2$}}
\put(36.00,19.50){\makebox(0,0)[lc]{$1'$}}
\put(36.00,9.50){\makebox(0,0)[lc]{$2'$}}
\put(36.00,-0.50){\makebox(0,0)[lc]{$3'$}}
\end{picture}\quad
\newpic{103}
\unitlength=0.50mm
\special{em:linewidth 0.6pt}
\linethickness{0.6pt}
\begin{picture}(46.00,1.00)(-4,11)
\emline{20.00}{19.50}{1}{35.00}{19.50}{2}
\emline{20.00}{19.50}{3}{5.00}{24.50}{4}
\emline{20.00}{19.50}{5}{5.00}{14.50}{6}
\emline{5.00}{4.50}{7}{20.00}{4.50}{8}
\emline{20.00}{4.50}{9}{35.00}{9.50}{10}
\emline{20.00}{4.50}{11}{35.00}{-0.50}{12}
\emline{20.00}{19.50}{13}{20.00}{4.50}{14}
\put(20.00,4.50){\circle*{2.00}}
\put(20.00,19.50){\circle*{2.00}}
\put(4.00,24.50){\makebox(0,0)[rc]{$3$}}
\put(4.00,14.50){\makebox(0,0)[rc]{$2$}}
\put(4.00,4.50){\makebox(0,0)[rc]{$1$}}
\put(36.00,19.50){\makebox(0,0)[lc]{$2'$}}
\put(36.00,9.50){\makebox(0,0)[lc]{$1'$}}
\put(36.00,-0.50){\makebox(0,0)[lc]{$3'$}}
\end{picture}
\ee

\bigskip\noindent
have the unpleasant peculiarity to go to infinity for the forward
scattering. For instance the first diagram from (44) is
\be
{1\over
m^2+(p_1+p_2-p'_2)^2}={1\over(p_2-p'_2)^2+2(p_2-p'_2)p_1}
{\to}\infty,\quad
{\rm when}\quad p_2\to p'_2.
\ee

Nevertheless the expression for $\vphantom{|^2}
\newpic{104}
\unitlength=1.00mm
\special{em:linewidth 0.4pt}
\linethickness{0.4pt}
\begin{picture}(20.00,1.00)
\put(10.00,1.00){\makebox(0,0)[cc]{$\sc T$}}
\put(16.00,1.00){\makebox(0,0)[cc]{$\sc T$}}
\put(4.00,1.00){\makebox(0,0)[cc]{$\sc T$}}
\put(16.00,1.00){\circle{6.00}}
\put(10.00,1.00){\circle{6.00}}
\put(4.00,1.00){\circle{6.00}}
\put(7.00,1.00){\circle*{1.00}}
\put(13.00,1.00){\circle*{1.00}}
\end{picture}
$
where the diagrams (44) give the
contribution is finite indeed:
because of
$$
\newpic{105}
\unitlength=1.00mm
\special{em:linewidth 0.4pt}
\linethickness{0.4pt}
\begin{picture}(8.00,1.00)
\put(1.00,1.00){\circle*{1.00}}
\put(7.00,1.00){\circle*{1.00}}
\put(4.00,1.00){\circle{6.00}}
\put(4.00,1.00){\makebox(0,0)[cc]{$\sc T$}}
\end{picture}
=-{\partial\over\partial m^2}
\newpic{106}
\unitlength=1.00mm
\special{em:linewidth 0.4pt}
\linethickness{0.4pt}
\begin{picture}(8.00,1.00)
\put(7.00,1.00){\circle*{1.00}}
\put(4.00,1.00){\circle{6.00}}
\put(4.00,1.00){\makebox(0,0)[cc]{$\sc T$}}
\end{picture}
={1\over(2\pi)^3}\int{d^3\bp\over
2\omega(e^{\beta\omega}-1)}\cdot{1\over\bp^2},
$$
one has
\be
3
\newpic{107}
\unitlength=1.00mm
\special{em:linewidth 0.4pt}
\linethickness{0.4pt}
\begin{picture}(20.00,1.00)
\put(10.00,1.00){\makebox(0,0)[cc]{$\sc T$}}
\put(16.00,1.00){\makebox(0,0)[cc]{$\sc T$}}
\put(4.00,1.00){\makebox(0,0)[cc]{$\sc T$}}
\put(16.00,1.00){\circle{6.00}}
\put(10.00,1.00){\circle{6.00}}
\put(4.00,1.00){\circle{6.00}}
\put(7.00,1.00){\circle*{1.00}}
\put(13.00,1.00){\circle*{1.00}}
\end{picture}
={4\over(2\pi)^9}\int\left(\prod_{k=1}^{3} {d^3\bp_k\over
2\omega_k(e^{\beta\omega_k}-1)}\right)\left({1\over\bp_1^2}+
{1\over\bp_2^2}+{1\over\bp_3^2}\right).
\ee
The way out of situation is to regularize the diagrams (44).
The recipe of appropriate regularization proceeds from comparison
(43), (45) and (46). Namely ``zero angle'' is to be understood as the
limit
$$
p'=\lim_{\varepsilon\to0}(p+\varepsilon), \quad
\varepsilon=(\varepsilon_0,\vec{\varepsilon}).
$$
The integration on phase volume is to be carried out at
$\varepsilon\neq0$ and put $\varepsilon=0$ in the final result.
Really, it is not hard to verify that in such limit the
integration of the first pare of diagrams (44) gives
\bea
&&\lim_{\vec{\varepsilon}\to0}\left[\lim_{\varepsilon_0\to0}\int
{d^3\bp_1\over 2\omega_1(e^{\beta\omega_1}-1)}\biggl({1\over
m^2+(p_1+\varepsilon)^2}+ {1\over
m^2+(p_1-\varepsilon)^2}\biggr)\right]=\nonumber\\
&&=\int {d^3\bp_1\over 2\omega_1(e^{\beta\omega_1}-1)}\,{1\over
\bp_1^2}\nonumber
\eea
in agreements with (46).

The problem with the diagrams (44) is particular case of more
general problem concerning $n\to n$ amplitudes which are built up from
rescattered blocks, such as
\be
\newpic{108}
\unitlength=0.75mm
\special{em:linewidth 0.4pt}
\linethickness{0.4pt}
\begin{picture}(55.00,11.00)
\put(4.00,0.00){\circle{6.00}}
\emline{1.00}{0.00}{1}{-5.00}{0.00}{2}
\emline{-5.00}{0.00}{3}{-5.00}{0.00}{4}
\emline{4.00}{3.00}{5}{4.00}{8.00}{6}
\emline{4.00}{-3.00}{7}{4.00}{-9.00}{8}
\put(16.00,0.00){\circle{6.00}}
\emline{13.00}{0.00}{9}{7.00}{0.00}{10}
\emline{7.00}{0.00}{11}{7.00}{0.00}{12}
\emline{16.00}{3.00}{13}{16.00}{8.00}{14}
\emline{16.00}{-3.00}{15}{16.00}{-9.00}{16}
\emline{42.00}{0.00}{17}{42.00}{0.00}{18}
\put(42.00,0.00){\circle{6.00}}
\emline{42.00}{3.00}{19}{42.00}{8.00}{20}
\emline{42.00}{-3.00}{21}{42.00}{-9.00}{22}
\emline{51.00}{0.00}{23}{45.00}{0.00}{24}
\emline{45.00}{0.00}{25}{45.00}{0.00}{26}
\emline{19.00}{0.00}{27}{23.00}{0.00}{28}
\emline{39.00}{0.00}{29}{35.00}{0.00}{30}
\put(4.00,11.00){\makebox(0,0)[cc]{$2'$}}
\put(16.00,11.00){\makebox(0,0)[cc]{$3'$}}
\put(29.00,11.00){\makebox(0,0)[cc]{$\cdots$}}
\put(42.00,11.00){\makebox(0,0)[cc]{$n'$}}
\put(-8.00,0.00){\makebox(0,0)[cc]{$1$}}
\put(29.00,0.00){\makebox(0,0)[cc]{$\cdots$}}
\put(54.00,0.00){\makebox(0,0)[cc]{$1'$}}
\put(4.00,-12.00){\makebox(0,0)[cc]{$2$}}
\put(16.00,-12.00){\makebox(0,0)[cc]{$3$}}
\put(29.00,-12.00){\makebox(0,0)[cc]{$\cdots$}}
\put(42.00,-12.00){\makebox(0,0)[cc]{$n$}}
\end{picture}
\ee
\

\bigskip
\noindent
The diagrams (47) is singular in physical region when incoming
momenta are equal to outcoming. The problem was intensively attacked
[5,6] and regularization recipes were proposed. As was shown $PT$
analyses gives the simple and natural way of solving this problem. It
must be noted also that due to the special structure of diagrams
including rescattered blocks the summation of subsets of such diagrams
(particular case is subset of so-called ring diagrams [1]) removes
zero angle divergences at all.

\section{Conclusion}

The remarkable simplicity and regularity of the representation (42),
(43) for the contribution of interaction to the free energy in the 2nd
order of $PT$ gives the base for natural generalization to higher
order of $PT$, namely
\be
f(T)-f_{\rm ideal}(T)=\sum_{n=2}^{\infty} f_n(T),
\ee
where
\be
f_n(T)=-{1\over n!}{1\over(2\pi)^{3n}}\int\left(\prod_{k=1}^{n}
{d^3\bp_k\over2\omega_k(e^{\beta\omega_k}-1)}\right){\rm Re}\,[
A_{n\to n}(p_1,\ldots,p_n;\, p_1,\ldots,p_n)].
\ee
Note that the representation for the temperature dependent part of 1PI
$\Gamma^{(2)}$-function of similar structure was recently obtained
[7].

In the representation (48), (49) is not hard to recognize the
so-called diagrams of the first type in the framework of $S$-matrix
formulation of statistical mechanics [8]. As for the so-called second
type diagrams we can say noting because they are constructed from
bilinear combinations of ${\rm Im} A$ and ${\rm Re} A$.  Corresponding
terms may appear in higher (at least 3rd) orders of $PT$.

The authors of [8] were generalized the Beth-Uhlenbeck formula and
derived in the framework of nonrelativistic quantum mechanics the
complete virial expansion. Being guided by invariant form of the
representation they have supposed its validity for the relativistic
case. Our calculations in the framework $\lambda\varphi^4$ QFT model
confirm their hypothesis with allowance some corrections like
$(n!)^{-1}$ factors in phase volume integrals reflecting the particles
identity. Also an ambiguity in regularization procedure for the zero
angle rescattering amplitudes is removed.

In our opinion the $S$-matrix formulation of statistical mechanics
would be good tool for the thermodynamic analysis of hadron matter
[9--12] due to huge amount experimental information and well
elaborated phenomenological models for the scattering amplitudes. So
we consider it important to verify the $S$-matrix formulation of
statistical mechanics by QFT methods.

There need be no doubt that a representation analogous to (48), (49)
can be derived for other renormalizable (at least normal, without
confinement) QFTs because the mentioned representation is rather
conditioned by general structure of $PT$ diagrams than by
peculiarities of interaction.  More over, proceeding from that the
thermodynamic values in this approach are expressed through the
functionals of $S$-matrix elements one can believe it being correct
also for hadron physics. A simple but forcible argument for this
follows from  the fact that Tr$(e^{-\beta\widehat{H}})$  is invariant
with respect to basis of state, so one can use in particular the set
of asymptotic, i.e.  hadronic states. Its completeness is guaranteed
by absence of any other except for hadrons observable asymptotic
states. Unfortunately the direct derivation of formulas like (42),
(43) from QCD, for example, is not possible because of the asymptotic
states there (and consequently the scattering amplitudes) can not be
perturbatively defined.

\medskip
The authors thanks to L.L.~Jenkovszky and E.S.Martynov for fruitful
discussions.

\medskip
The work is partially supported by grant 2.2/225 of the Ukrainian
Foundation of Fundamental Research and by grant INTAS-93-1038.

\newpage

{\bf Fig.\,1}
{\bf Fig.\,2}
{\bf Fig.\,3}
{\bf Fig.\,4}
{\bf Fig.\,5}
{\bf Fig.\,6}

\end{document}